\documentclass{llncs}

\usepackage{ifpdf}
\usepackage{color}               
\ifpdf
\usepackage{ae,aecompl}
\usepackage{graphicx}            
\else
\usepackage[dvips]{graphicx}
\DeclareGraphicsExtensions{.ps,.jpg,.jpeg,.eps,.pdf,.png,.mps}
\fi
\usepackage{picins}
\makeatletter
\def\piccaption{\@dblarg{\@piccaption}} 
\makeatother
\usepackage{amsmath,amssymb}
\usepackage{algorithm}
\usepackage{algorithmic}
\usepackage[latin1]{inputenc}
\usepackage[OT1]{fontenc}
\usepackage{comment}

\usepackage[all]{xy}

\newtheorem{theo}{Theorem}

\newenvironment{myproof}[1]{
\vspace{-5ex}

\hspace*{1em}\begin{trivlist}\item[\hskip\labelsep{\textsc{#1}}]}%
{\end{trivlist}}
\renewenvironment{proof}{
\begin{myproof}{Proof.}}{\hfill $\Box$\end{myproof}}

\newcommand{\IDEO}{\ensuremath{\call{I}_{\mbox{\tiny\rm DEO}}}}
\newcommand{\IIDEO}{\ensuremath{\call{I}'_{\mbox{\tiny\rm DEO}}}}
\newcommand{\call}[1]{\ensuremath{{\mathcal{#1}}}}

\newcommand{\mathconstante}[1]{\ensuremath{{\mathrm{#1}}}}

\newcommand{\Variables}{\call{X}}
\newcommand{\Constants}{\mathconstante{C}}
\newcommand{\fonction}[2]{\ensuremath{{\mathconstante{#1}(#2)}}}

\newcommand{\set}[1]{\ensuremath{{\left\lbrace #1 \right\rbrace}}}

\newcommand{\termset}[1]{{\fonction{T}{#1}}}

\newcommand{\rhnorm}[2]{\ensuremath{{(#1)\!\!\downarrow_{#2}}}}

\newcommand{\norm}[1]{\rhnorm{#1}{}}

\newcommand{\sub}[1]{\fonction{\text{Sub}}{#1}}

\newcommand{\Var}[1]{\fonction{Var}{#1}}
\newcommand{\Sub}[1]{\fonction{Sub}{#1}}

\newcommand{\Supp}[1]{\fonction{Supp}{#1}}

\newcommand{\grule}{\mathconstante{L}}
\newcommand{\lrule}[1]{{\ensuremath{\grule^{#1}}}}

\newcommand{\rhclos}[2]{{\ensuremath{\overline{#1}^{#2}}}}
\newcommand{\clos}[1]{\rhclos{#1}{}}

\newenvironment{decisionproblem}[1]{
\vspace*{-1ex}
\begin{tabbing}
  \underline{\textbf{#1}}\\
  \hspace*{2em}\= \textbf{Input:}~ \=}{
\end{tabbing}
\vspace*{-1em}
}

\newcommand{\change}{}
\newcommand{\entreeu}[1]{ \hbox{\vbox{\parbox[t]{0.8\linewidth}{#1}}}\\}
  
\newcommand{\sortie}[1]{
\> \textbf{Output:} \> \hbox{\vbox{\parbox[t]{0.8\linewidth}{#1}}}\\}

\newcommand{\II}{\ensuremath{\call{I}'_{\mbox{\tiny\rm }}}}
\newcommand{\I}{\ensuremath{\call{I}_{\mbox{\tiny\rm }}}}
\newcommand{\unif}{\ensuremath{\stackrel{?}{=}}}
\newcommand{\intrus}[3]{\ensuremath{\left\langle #1,#2,#3\right\rangle}}
\newcommand\PP[1]{\fonction{P}{#1}}
\newcommand{\IIDSKS}{\ensuremath{\call{I}'_{\mbox{\tiny\rm DSKS}}}}
\newcommand{\nnintrus}[1]{\intrus{\call{G}}{S}{\emptyset}}
\newcommand{\IDSKS}{\ensuremath{\call{I}_{\mbox{\tiny\rm DSKS}}}}
\newcommand{\nintrus}[1]{\intrus{\call{G}}{S}{\call{H}}}

\newcommand{\gsig}[1]{\termset{#1}}
\newcommand{\vsig}[1]{\termset{#1,\Variables}}

\newcommand{\TGX}{\vsig{\call{G}}}

\newcounter{lemc}
\newcommand{\rew}{\rightarrow}
\newcommand{\nar}{\leadsto}
\newcommand{\bnar}{\leadsto_{\text{\rm b.n.}}}
\newcommand{\ded}{\twoheadrightarrow}

\newcommand{\arity}[1]{\ensuremath{\mathop{\mbox{\rm ar}}(#1)}}

\newcommand{\M}[1]{\ensuremath{\mathop{\mbox{\rm M}}(#1)}}

\newcommand{\nbv}[1]{\ensuremath{\mathop{\mbox{\rm nbv}}(#1)}}

\newcommand{\SK}[1]{\ensuremath{\mathop{\mbox{\rm SK}}(#1)}}
\newcommand{\PK}[1]{\ensuremath{\mathop{\mbox{\rm PK}}(#1)}}
\newcommand{\SKp}[2]{\ensuremath{\mathop{\mbox{\rm S'K}}(#1,#2)}}
\newcommand{\PKp}[2]{\ensuremath{\mathop{\mbox{\rm P'K}}(#1,#2)}}
\newcommand{\SSKp}[2]{\ensuremath{\mathop{\mbox{\rm S"K}}(#1,#2)}}
\newcommand{\PPKp}[2]{\ensuremath{\mathop{\mbox{\rm P"K}}(#1,#2)}}
\newcommand{\Sig}[2]{\ensuremath{\mathop{\mbox{\rm Sig}}(#1,#2)}}
\newcommand{\Ver}[3]{\ensuremath{\mathop{\mbox{\rm Ver}}(#1,#2,#3)}}
\newcommand{\f}[2]{\ensuremath{\mathop{\mbox{\rm f}}(#1,#2)}}

\title{Key Substitution in the Symbolic Analysis of Cryptographic
  Protocols (extended version)}

\author{Yannick Chevalier \and Mounira Kourjieh}
\institute{IRIT, Universit{\'e} de Toulouse, France  \\
  email: $\lbrace$ychevali,kourjieh$\rbrace$@irit.fr }

\usepackage{proof}
\begin{document}
\sloppy

\maketitle{}

\begin{abstract}
   Key substitution vulnerable signature schemes are signature schemes
  that permit an intruder, given a public verification key and a
  signed message, to compute a pair of signature and verification
  keys such that the message appears to be signed with the new
  signature key.  Schemes vulnerable to this attack thus permit an
  active intruder to claim to be the issuer of a signed message. 
  
A digital signature scheme is said to be vulnerable to \emph{ destructive exclusive ownership property (DEO)}
If it is computationaly feasible for an intruder, given a public verification key and  a pair of message and its valid signature relatively to the given public key $(m,s)$, to compute a pair of signature and verification keys and a new message $m'$ such that  $s$ is a valid signature of $m'$ relatively to the new verification key.

In  this paper, we investigate and solve positively the problem of the
  decidability of symbolic cryptographic protocol analysis when the
  signature schemes employed in the concrete realisation have this two properties. 
\end{abstract}

\section{Introduction}


According to
\emph{West's Encyclopedia of American Law}, a \emph{signature} is
\begin{quote}
  ``A mark or sign made by an individual on an instrument or document
  to signify knowledge, approval, acceptance, or obligation\ldots [Its
  purpose] is to authenticate a writing, or \emph{provide notice of
    its source}\footnote{We have emphasised}\ldots''
\end{quote}
We will not deal any further with legal considerations, but it is
interesting to note that while digital signatures are primarily
employed to authenticate a document, \textit{i.e.} ensure that the
signer endorses the content of the document, they can also be employed
to prove the origin of a document, \textit{i.e.} ensure that only one
person could have signed it.  Indeed, most of the cryptographic work
on digital signatures has aimed at certifying that no-one could sign a
document in the place of someone else.

The analysis of digital signature primitives has however focused on
the former authentication property.  Formally speaking, the yardstick
security notion for assessing the robustness of a digital signature
scheme is the existential enforceability against adaptative
chosen-message attacks (UNF-CCA)~\cite{GoldwasserMR88}. This notion
states that, given a signing key/verification key pair, it is
infeasible for someone ignorant of the signing key to forge a message
that can pass the verification with the public verification key, and
this even when messages devised by the attacker are signed beforehand.
The security goal provided by this property is the impossibility
(within given computing bounds) to impersonate a legitimate user
(\emph{i.e.}  one that does not reveal its signature key) when signing
a message.

We note that this robustness does not address the issue of the
identification of a source of a message. However, this latter concept
is also pertaining to digital signatures when they are employed in a
non-repudiation protocol. While one would not differentiate the two
properties at first glance, they are different since the
authentication property requires the existence of the participation of
the signer in the creation of the message, while the latter mandates
the unicity of a possible creator of a message. 

The two notions of message authentication and source authentication
collapse in the \emph{single-user} setting when there exists only one
pair of signature/verification keys. They may however be different
in a \emph{multi-user} setting. We believe that the first work in this
direction was the discovery of a flaw on the Station-to-Station
protocol by Blake-Wilson and Menezes~\cite{Blake-WilsonM99}, where the
authors show how it is possible to confuse a participant into thinking
it shares a key with another person than the actual one. The attack
consisted in the creation, by the attacker, of a
signature/verification key pair dependent upon messages sent in the
protocol. Defining a signature scheme to have the Duplicate Signature
Key Selection (DSKS) property if it permits such a construction with
non-negligible probability, they showed that several standard
signature schemes (including RSA, DSA, ECDSA and ElGamal) had this
property, but also that a simple counter-measure (signing the public
key along with the message) existed in all cases, but was rarely
implemented. This DSKS property was formally defined as \emph{Key
  substitution} in~\cite{MenezesS04}, where it is also discussed,
after a review of what could be called an attack on a signature scheme
in the multi-user setting. It was also later presented independently
in~\cite{PorninS05} as \emph{Conservative Exclusive Ownership}. The
companion property of \emph{Destructive Exclusive Ownership} by which
an intruder may also change arbitrarily the signed message is also
introduced and they showed that the usual signature algorithms (such as RSA and DSS)  have this property.
While the same attacks as in ~\cite{MenezesS04} are
exhibited, the authors also demonstrate how this can be used in
practice to poison a badly implemented PKI with fake CRLs (T. Pornin, 
personal communication).

\paragraph{Automated validation of security protocols.} 
Cryptographic protocols have been applied to securing communications
over an insecure network for many years. While these protocols rely on
the robustness of the employed security primitives, their design is
error-prone. This difficulty is reflected by the repeated discovery of
logical flaws in proposed protocols, even under the assumption that
cryptographic primitives were perfect.  As an attempt to solve the
problem, there has been a sustained effort to devise formal methods
for specifying and verifying the security goals of protocols. Various
symbolic approaches have been proposed to represent protocols and
reason about them, and to attempt to verify security properties such
as confidentiality and authenticity, or to discover bugs. Such
approaches include process algebra, model-checking, 
equational reasoning, constraint solving and resolution
theorem-proving (e.g.,
~\cite{Weidenbach99,amadio2003,ChevalierV-IEEE-ASE01,ArmandoCompagna02b}).

Our goal is to adapt the symbolic model of concrete cryptographic
primitives in order to reflect inasmuch as possible their
imperfections that could be used by an attacker to find a flaw on a
protocol. The work described in this paper relies on the
compositionality result obtained in~\cite{ChevalierR05} that permits
us to abstract from other primitives and consider protocols that only
involve a signature scheme having DSKS property (resp. vulnerable to DEO property).

\paragraph{Outline.} In Section~\ref{sec:attack} we will present an
attack by Baek \emph{et al.} demonstrating how an actual intruder can
use the DSKS property of a signature scheme to attack a protocol. We
then describe in Section~\ref{sec:setting} the formalism in which we
will analyse cryptographic protocols. In Section~\ref{sec:dsks} we
present how we model the possible actions of an intruder taking
advantage of the DSKS property of a signature scheme and in section~\ref{sec:DEO}, we present how we model the possible actions of an intruder taking advantage of the vulnerability of signature scheme to DEO property. We present in
Section~\ref{sec:saturation} an algorithm that permits to reduce the
analysis to an analysis in the empty equational theory, and give in
Section~\ref{sec:decision} a decision procedure for the reachability
problem in these protocols. We conclude in
Section~\ref{sec:conclusion}.


\section{An example of attack\label{sec:attack}}
We do not present here the original attack on the station-to-station
protocol, but one that we believe to be simpler, and given by
\emph{Baek et al.}~\cite{SCIS2000} on the KAP-HY (\emph{Key Agreement
  protocol}, proposed by \emph{Hirosi} and \emph{Yoshida}
in~\cite{HiroseY98}).

\paragraph{Presentation of the KAP-HY protocol.} This protocol relies
on a redundant signature scheme to provide key confirmation at the end
of a key exchange. The signature of a message $m$ by agent $A$ is
denoted $s_A(m)$. Abstracting the details of the Diffie-Hellman key
construction with messages $u_A$ and $u_B$, and of the signature
scheme, the protocol reads as follows:\\
$~~~~~~~~~~~~~~~~~~~~~~~~\begin{array}{rl}
  A\to B:& u_A,A\\
  B\to A:& u_B,s_B(u_A),B\\
  A\to B:& s_A(s_B(u_A),u_B)\\
\end{array}
$ 

An unknown key share (UKS) attack on a key agreement protocol is an
attack whereby two entities $A$ and $B$ participating in a key
agreement protocol may end the protocol successfully, but with a wrong
belief on who shares a key with who.  In \cite{SCIS2000}, \emph{Baek
  et al.} showed that the redundant signature scheme employed in the
KAP-HY protocol possesses the DSKS property, and elaborate on this to
show that the KAP-HY is vulnerable to a UKS attack. In this attack, the intruder $E$ waits that $A$ initiates a session with him:
\\
$~~\begin{array}{lrl@{\hspace*{6em}}lrl}

(1) & A\to E:& u_A,A  &
(2)&  E\to A:& u_B,s_B(u_A),E\\
(1')&  E\to B:&  u_A,A &
(3) & A\to E:& s_A(s_B(u_A),u_B)\\
(2') & B\to E(A):&  u_B,s_B(u_A),B &
(3')&  E\to B:&  s_A(s_B(u_A),u_B)\\
\end{array}
$

In this attack, the intruder $E$ records, but passes unchanged, the
first message, and initiates a session as $A$ with $B$. It then
intercepts the second message, and builds from the public key of $B$
and from the message $s_B(u_A)$ a signature/verification key pair, and
registers this key pair. $E$ then passes the signature, but this time
accompanied by its identity (2').  The main point is that when $A$
checks the signature of the incoming message, it accepts it on the
ground that it seems to originate from $E$.  At the end of this
execution, $A$ believes that the key is shared with $E$ whereas it is
actually shared with $B$.

The computation of the new pair of keys $(P_E,S_E)$ proceeds as
follows.  At the end of flow (2), the intruder knows the signature of
$u_A$ made by Bob using his public key, then, by using DSKS property
of the used signature scheme, he creates the new pair of keys
$(P_E,S_E)$. The crucial point, common to all DSKS attacks, is the
construction of a new key pair from a public verification key and from
a signed message. We will model this operation with appropriate
deduction rules, and prove that protocol analysis remains decidable.

\section{Formal setting} \label{sec:setting}

\subsection{Basic notions\label{subsec:setting:notions}}

We consider an infinite set of free constants \Constants{} and an
infinite set of variables \Variables.  For any signature \call{G}
(\textit{i.e.}  sets of function symbols not in \Constants{} with arities) we
denote \gsig{\call{G}} (resp.  \vsig{\call{G}}) the set of terms over
$\call{G}\cup{}\Constants{}\change{}$ (resp.
$\call{G}\cup{}\Constants{}\change{}\cup\Variables$).  The former is
called the set of ground terms over \call{G}, while the latter is
simply called the set of terms over \call{G}.  The arity of a function
symbol $g$ is denoted by \arity{g}.  Variables are denoted by $x$,
$y$, terms are denoted by $s$, $t$, $u$, $v$, and finite sets of terms
are written $E,F,...$, and decorations thereof, respectively.  We
abbreviate $E\cup F$ by $E,F$, the union $E\cup\{t\}$ by $E,t$ and
$E\setminus \{t\}$ by $E\setminus t$.  The \emph{subterms} of a term
$t$ are denoted $\sub{t}$ and are defined recursively as follows. If
$t$ is an atom (\textit{i.e.}  $t\in \Variables\cup \Constants{}$)
then $\sub{t}=\set{t}$. If $t=g(t_1,\ldots,t_n)$ then
$\sub{t}=\set{t}\cup\bigcup_{i=1}^n\sub{t_i}$. The \emph{positions} in
a term $t$ are sequences of integers defined recursively as follows,
$\varepsilon$ being the empty sequence representing the root position
in $t$.  We write $p\leq q $ to denote that the position $p$ is a
prefix of position q.  If $u$ is a subterm of $t$ at position $p$ and
if $u=g(u_1,\ldots,u_n)$ then $u_i$ is at position $p\cdot{}i$ in $t$
for $i\in\set{1,\ldots,n}$.  We write $t_{|p}$ the subterm of $t$ at
position $p$. We denote $t[s]$ a term $t$ that admits $s$ as subterm.
The size $\|t\|$ of a term $t$ is the number of distinct subterms of
$t$.  The notation is extended as expected to a set of terms.

A substitution $\sigma$ is an involutive mapping from \Variables{} to
\vsig{\call{G}} such that $\Supp{\sigma}=\{ x|\sigma(x)\not=x\}$, the
\emph{support} of $\sigma$, is a finite set.  The application of a
substitution $\sigma$ to a term $t$ (resp.  a set of terms $E$) is
denoted $t\sigma$ (resp. $E\sigma$) and is equal to the term $t$
(resp. $E$) where all variables $x$ have been replaced by the term
$\sigma(x)$. A substitution $\sigma $ is \emph{ground} w.r.t.
$\call{G}$ if the image of $\Supp{\sigma }$ is included in
$\gsig{\call{G}}$.

An \emph{equational presentation} $\call{H}=(\call{G},\call{A})$ is defined
by a set \call{A} of equations $u=v$ with $u,v\in\TGX{}$ and $u,v$ without
free constants. For any equational presentation \call{H} the relation
$=_{\call{H}}$ denotes the equational theory \emph{generated} by
$(\call{G},\call{A})$ on \TGX{}, that is the smallest congruence containing
all instances of axioms of $A$. Abusively we shall not distinguish
between an equational presentation \call{H} over a signature \call{G}
and a set \call{A} of equations presenting it and we denote both by
\call{H}. If the equations of \call{A} can be oriented from left to
right, we write the equations in \call{A} with an arrow, $l\rew r$.
The equations can then only be employed from left to right, and
\call{A} is called a rewrite system. An equational theory can in this
case be defined by a rewrite system.  An equational theory \call{H} is
said to be \emph{consistent} if two free constants are not equal
modulo \call{H} or, equivalently, if it has a model with more than one
element modulo \call{H}.

Let \call{A} be a set of rewrite rules $l\rew r$.  The rewriting
relation $\rew_\call{A}$ between terms is defined by $t\rew_\call{A}
t'$ if there exists $l\to r\in \call{A}$ and a substitution $\sigma$
such that $l\sigma=s$ and $r\sigma=s'$, $t=t[s]$ and $t'=t[s\leftarrow
s']$.  \call{A} is convergent if and only if it is terminating and
confluent.  In this case, all rewriting sequences starting from $t$
are finite and have the same limit, and this limit is called the
\emph{normal form} of $t$.  We denote this normal form
$\norm{t}_{\call{A}}$, or \norm{t} when the considered rewriting
system is clear from the context.  A substitution $\sigma$ is in
normal form if for all $x\in\Supp{\sigma}$, the term $\sigma(x)$ is in
normal form.

\subsection{Unification systems}

In the rest of this section, we let $\call{H}$ be an equational theory
on \TGX{} and \call{A} be a \emph{convergent} rewriting system
generating $\call{H}$.

\begin{definition}{\label{def:unification}(Unification systems)}
  Let $\call{H}$ be an equational theory on \TGX.  A
  $\call{H}$-\emph{unification system} \call{S} is a finite set of
  pairs of terms in \TGX{} denoted by $\{u_i \unif{}_{\call{H}}
  v_i\}_{i\in\set{1,\ldots,n}}$.  It is satisfied by a substitution
  $\sigma$, and we note $\sigma\models{}_{\call H}\call{S}$, if for
  all $i\in\set{1,\ldots,n}$ we have $u_i\sigma =_\call{H} v_i\sigma$.
  In this case we call $\sigma$ a \emph{solution} or a \emph{unifier}
  of \call{S}.
\end{definition}

When \call{H} is generated by \call{A}, the confluence implies that if
$\sigma$ is a solution of a \call{H}-unification system, then
\norm{\sigma} is also a solution of the same unification system.
Accordingly we will consider in this paper only solutions in normal
form of unification systems. A \emph{complete set of unifiers} of a
\call{H}-unification system \call{S} is a set $\Sigma$ of
substitutions such that, for any solution $\tau$ of \call{S}, there
exists $\sigma\in\Sigma$ and a substitution $\tau'$ such that
$\tau=_\call{H}\sigma\tau'$. The unifier $\tau$ is a \emph{most
  general unifier} of \call{S} if the substitution $\tau'$ in the
preceding equation must be a variable renaming.

In the context of unification modulo an equational theory, standard
(or syntactic) unification will also be called unification in the
empty theory. In this case, it is well-known that there exists a
unique most general unifier of a set of equations. This unifier is
denoted $mgu(\call{S})$, or $mgu(s,t)$ in the case
$\call{S}=\set{s\unif_\emptyset t}$.

\begin{decisionproblem}{Unifiability Problem}
  \entreeu{ A \call{H}-unification system \call{S}.}%
  \sortie{\textsc{Sat} iff there exists a substitution $\sigma$ such
    that $\sigma\models_{\call{H}}\call{S}$.}
\end{decisionproblem}

Let us now introduce the notion of narrowing, that informally permits
to instantiate and rewrite a term in a single step.

\begin{definition}{\label{def:narrowing}(Narrowing)}
  Let $s$ and $t$ be two terms.  We say $t \nar s$ iff there
  exists $l\rew r \in \mathcal{A}$, a position $p$ such that
  $t_{|p}\notin\Variables$
  and $s= t\sigma [p \leftarrow r\sigma],$ where
  $\sigma=mgu(t_{|p},l)$.  We denote by $\nar$ the narrowing
  relation.
\end{definition}
Assume $t\leadsto t'$ with a rule $l\to r$ applied at a position $p$ in $t$.
A basic position in $t'$ is either a non-variable position of $t$ not under $p$ or a position $p\cdot q$
where $q$ is a non-variable position in $r$. Basic narrowing is a restricted form of narrowing where only terms at basic positions are considered to be narrowed.  In the rest of this paper,  we  denote
$t\bnar t'$ a basic narrowing step.

\subsection{Intruder deduction systems}

The notions that we give here have been defined
in~\cite{ChevalierR05}.  These definitions have since been generalised
to consider a wider class of intruder deduction systems and constraint
systems~\cite{frocos07}. Although this general class encompasses all
intruder deduction systems and constraint systems given in this paper,
we have preferred to give the simpler definitions
from~\cite{ChevalierR05} which are sufficient for stating our problem.
We will refer, without further justifications, to the model
of~\cite{frocos07} as \emph{extended} intruder systems and
\emph{extended} constraint systems. The latter correspond to symbolic
derivations in which a most general unifier of the unification system
has been applied on the input/output messages.

In the context of a security protocol (see
\textit{e.g.}~\cite{meadows96} for a brief overview), we model
messages as ground terms and intruder deduction rules as rewrite rules
on sets of messages representing the knowledge of an intruder.  The
intruder derives new messages from a given (finite) set of messages by
applying deduction rules.  Since we assume some equational axioms
$\call{H}$ are satisfied by the function symbols in the signature, all
these derivations have to be considered \emph{modulo} the equational
congruence $=_{\call{H}}$ generated by these axioms.  In the setting
of~\cite{ChevalierR05} an intruder deduction rule is specified by a
term $t$ in some signature \call{G}.  Given values for the variables
of $t$ the intruder is able to generate the corresponding instance of
$t$.

\begin{definition}{\label{def:intruder}}
  An \emph{intruder system} $\call I$ is given by a triple
  \intrus{\call{G}}{S}{\call{H}} where $\call{G}$ is a
  signature, $S \subseteq \TGX$ and $\call H$ is a set of equations
  between terms in $\TGX $.  To each $t\in S$ we associate a
  \emph{deduction rule} $\lrule{t}:\Var{t}\ded{}t$ .   
  The set of rules $\grule{}_{\call I}$ is defined as the union 
  of $\lrule{t}$ for  all $t\in S$.
\end{definition}

Each rule $l\ded{}r$ in $\grule{}_{\call I}$ defines an intruder
deduction relation $\ded_{l\ded{}r}$ between finite sets of terms.
Given two finite sets of terms $E$ and $F$ we define
$E\ded_{l\ded{}r}F$ if and only if there exits a substitution
$\sigma$, such that $l\sigma=_{\call{H}}l'$, $r\sigma=_{\call{H}}r'$,
$l'\subseteq{}E$ and $F = E \cup \set{r'}$.  We denote $\ded_{\call
  I}$ the union of the relations $\ded_{l\ded{}r}$ for all $l\ded{}r$
in $L_{\call I}$ and by $\ded_{\call I}^*$ the transitive closure of
$\ded_{\call I}$.  Note that by definition, given sets of terms $E$,
$E'$, $F$ and $F'$ such that $E=_\call{H}E'$ and $F=_\call{H}F'$ by
definition we have $E\ded_{\call I}F$ iff $E'\ded_{\call I}F'$. We
simply denote by $\ded$ the relation $\ded_{\call I}$ when there is no
ambiguity about ${\call I}$.

A \emph{derivation} $D$ of length $n$, $n\ge 0$, is a sequence of
steps of the form $E_0\ded_{\call I} E_0,t_1\ded_{\call
  I}\cdots\ded_{\call I} E_n$ with finite sets of terms
$E_0,\ldots{}E_n$, and terms $t_1,\ldots,t_n$, such that
$E_i=E_{i-1}\cup{}\{t_i\}$ for every $i\in \set{1,\ldots,n}$.  The
term $t_n$ is called the {\em goal} of the derivation. We define
$\rhclos{E}{\call I}$ to be equal to the set of terms that can be
derived from $E$. If there is no ambiguity on the intruder deduction
system $\call I$ we write \clos{E} instead of $\rhclos{E}{\call I}$.

\subsection{Simultaneous constraint satisfaction problems\label{reachprob}} 

We now introduce the constraint systems to be solved for checking
protocols. It is presented in~\cite{ChevalierR05} how these constraint
systems permit to express the reachability of a state in a protocol
execution.

\begin{definition}{\label{def:constraints}(\call{I}-Constraint systems)} 
  Let ${\call I} =\langle \call{G}, S, \call{H} \rangle $ be an
  intruder system.  An $\call I$-\emph{constraint system} \call{C} is
  denoted $((E_i\rhd{}v_i)_{i\in\set{1,\ldots,n}},\call{S})$ and is
  defined by a sequence of pairs $(E_i, v_i)_{i\in\set{1,\ldots,n}}$
  with $v_i\in\Variables{}$, $E_i \subseteq \TGX$ for
  $i\in\set{1,\ldots,n}$, and $E_{i-1}\subseteq E_{i}$ for
  $i\in\set{2,\ldots,n}$, and
  $\Var{E_i}\subseteq\set{v_1,\ldots,v_{i-1}}$ and by an $\call
  H$-unification system \call{S}.
  
  An $\call I$-\emph{Constraint system} \call{C} is satisfied by a
  substitution $\sigma$ if for all $i\in\set{1,\ldots,n}$ we have
  $v_i\sigma\in\rhclos{E_i\sigma}{\call{I}}$ and if
  $\sigma\models_\call{H}{}\call{S}$. We denote that a substitution
  $\sigma$ satisfies a constraint system \call{C} by
  $\sigma\models_{\call I}\call{C}$.
\end{definition}

Constraint systems are denoted by \call{C} and decorations thereof.
Note that if a substitution $\sigma$ is a solution of a constraint
system \call{C}, by definition of deduction rules and unification
systems the substitution \norm{\sigma} is also a solution of \call{C}.
In the context of cryptographic protocols the inclusion
$E_{i-1}\subseteq E_{i}$ means that the knowledge of an intruder does
not decrease as the protocol progresses: after receiving a message a
honest agent will respond to it, this response can then be added to
the knowledge of the intruder who listens to all communications.  The
condition on variables stems from the fact that a message sent at some
step $i$ must be built from previously received messages recorded in
the variables $v_j, j<i $, and from the ground initial knowledge of
the honest agents.  

Our goal will be to solve the following decision problem for the
intruder deduction system modelling a signature scheme having the DSKS
property.

\begin{decisionproblem}{\call{I}-Reachability Problem}
  \entreeu{ An \call{I}-constraint system \call{C}.}
  \sortie{\textsc{Sat} iff there exists a substitution $\sigma$ such
    that $\sigma\models_{\call I}\call{C}.$}
\end{decisionproblem}

\section{Symbolic model for key substitution attacks}\label{sec:dsks}
A digital signature scheme is defined by three algorithms: the
signing algorithm, the verification algorithm and the key generation
algorithm. The last algorithm generates for each user a pair of keys,
one of them will be used as signing key and will be kept secret, while
the other is public and will be used as a verifying key. We abstract
the key generation algorithm with two functions, $\PK{\_}$ and
$\SK{\_}$ denoting respectively the verification and signature keys of
an agent.  We assume it is not possible, given an agent's name $A$, to
\emph{compute} $\PK{A}$ or $\SK{A}$. The signature of a message $m$
with signature key $k$ is a public algorithm $\Sig{\_}{\_}$, and the
resulting signed message is denoted $\Sig{m}{k}$. We consider
signatures with appendix, where the verification algorithm
$\Ver\_\_\_$ --which is available to everyone-- takes in its input a
message $m$, a signature $s$ and the public verification key $k$. The
application of the algorithm is denoted $\Ver{m}{s}{k}$, and its
outcome can be $0$ ($s$ is not the signature of $m$ with the signature
key associated with the verification key $k$) or $1$ ($s$ is a valid
signature).

In addition to these functions, we add two new functions, $\PKp\_\_$ and
$\SKp\_\_$, which are public and take as argument a signed message $s$ and 
a verification key $k$ corresponding to this signed message, and output
respectively a verification and a signature key denoted $\PKp{s}{k}$ and
$\SKp{s}{k}$. The verification of $s$ with the verification key
$\PKp{s}{k}$ succeeds. 

Given this informal description, the equational theory \call{H_{DSKS}}
to which these operations abide by is presented by the following set
\call{A_{DSKS}} of equations:

$$
\call{A_{DSKS}}= 
\left\lbrace
  \begin{array}[c]{l}
    \Ver{x}{\Sig{x}{\SK{y}}}{\PK{y}}=1\\
    \Ver{x}{\Sig{x}{\SKp{y_1}{y_2}}}{\PKp{y_1}{y_2}}=1\\
    \Sig{x} { \SKp{\PK{y}}{\Sig{x}{\SK{y}} }  }= \Sig{x}{\SK{y}}\\
  \end{array}
\right.
$$

The public operations defined above are now translated into an
intruder system
$\IDSKS{}=\left\langle\call{G_{DSKS}},\call{L_{DSKS}},\call{H_{DSKS}}\right\rangle$
with:
$$
\left\lbrace
  \begin{array}{rcl}
    \call{G_{DSKS}}&=&\set{\text{Sig}, \text{Ver},  \SKp\_\_, \PKp\_\_ , 0 ,1,\text{SK},\text{PK}}\\
    \call{L_{DSKS}}&=&  \set {\Sig{x}{y}, \Ver{x}{y}{z}, \SKp{x}{y}, 
      \PKp{x}{y}, 0, 1}\\
  \end{array}
\right.
$$

Note that the presentation \call{A_{DSKS}} is not convergent, and thus
we cannot apply results on basic narrowing as is. To this end we
introduce a rewriting system \call{R_{DSKS}} which is convergent and obtained
by Knuth-Bendix~\cite{KB} completion on \call{A_{DSKS}}, and such that
two terms have the same normal form for \call{R_{DSKS}} iff they are equal
modulo \call{H_{DSKS}}.

\begin{lemma}{\label{lemma:00:convergent}}
  \call{H_{DSKS}} is generated by the convergent rewriting system:
  $$
  \call{R_{DSKS}}=\left\lbrace
    \begin{array}[c]{l}
      \Ver{x} {\Sig{x}{\SK{y}}} {\PK{y}} \to 1\\
      \Ver{x} {\Sig{x}{\SKp{y_1}{y_2}}} {\PKp{y_1}{y_2}} \to 1\\
      \Ver{x}{\Sig{x}{\SK{y}} }  {\PKp{\PK{y}}{\Sig{x}{\SK{y}} }} \to 1 \\
      \Sig{x}{ \SKp{\PK{y}}{\Sig{x}{\SK{y}} }  } \to \Sig{x}{\SK{y}} 
    \end{array}
  \right.
  $$
\end{lemma}
\begin{proof}
The application of  the \emph{Knuth-Bendix} completion
procedure~\cite{KB} to $\call{H_{DSKS}}$ gives us the convergent rewriting system 
$\mathcal{R_{DSKS}}$. This rewriting system generates $\call{H}_{DSKS}$, this conclude the proof.

\end{proof}

\section{Symbolic model for DEO attacks}\label{sec:DEO}
A digital signature scheme is vulnerable against  \emph{destructive exclusive ownership (DEO)} if it is computationaly feasible for the intruder, given $K_{pub}$ and a pair $(m,s)$ such that $\Ver{K_{pub}}{m}{s}=1$, to produce values $K'_{pub},~ K'_{priv},~ m'$ and $s'$ such that 
$K'_{pub}\not= K_{pub}$, $K'_{priv}$ matches $K'_{pub}$, $s'=s$, $m'\not= m$ and $\Ver{K'_{pub}}{m'}{s}=1$.

A digital signature scheme is defined by three algorithms: the signing algorithm, the verification algorithm and the key generation algorithm. These algorithms  have the same properties as above (section~\ref{sec:dsks}). We abstract the signature scheme with the following functions symbols: $\PK{\_}$, $\SK{\_}$, $\Sig{\_}{\_}$ and $\Ver\_\_\_$.
In order to model DEO attacks, we introduce three functions symbols, $\PPKp\_\_$, $\SSKp\_\_$ and $\f\_\_$ which are public and
take as argument a signed message $s$ and 
a public verification key $k_{pub}$ corresponding to this signed message, and output
respectively a new verification key, a new signature key  and a new message denoted $\PPKp{s}{k}$,
$\SSKp{s}{k_{pub}}$ and  $\f{s}{k_{pub}}$. The verification of $s$ with the new public key
$\PPKp{s}{k_{pub}}$ and the message $\f{s}{k_{pub}}$ succeeds.

Given this informal description, the equational theory \call{H_{DEO}}
to which these operations abide by is presented by the following set
\call{A_{DEO}} of equations:

$$
\call{A_{DEO}}= 
\left\lbrace
  \begin{array}[c]{l}
    \Ver{x}{\Sig{x}{\SK{y}}}{\PK{y}}=1\\
    \Ver{x}{\Sig{x}{\SSKp {y_1}{y_2} }}{\PPKp {y_1}{y_2} }=1\\
    \Sig{\f{\PK{y}}{\Sig{x}{\SK{y}}}} { \SSKp    {\PK{y}}{\Sig{x}{\SK{y}} }  }= \Sig{x}{\SK{y}}\\
  \end{array}
\right.
$$

The public operations defined above are now translated into an
intruder system
$\IDEO{}=\left\langle\call{G_{DEO}},\call{L_{DEO}},\call{H_{DEO}}\right\rangle$
with:
$$
\left\lbrace
  \begin{array}{rcl}
    \call{G_{DEO}}&=&\set{\text{Sig}, \text{Ver},  \SSKp\_\_, \PPKp\_\_ , \text{f},  0, 1, \text{SK},\text{PK}}\\
    \call{L_{DEO}}&=&  \set {\Sig{x}{y}, \Ver{x}{y}{z}, \SSKp{x}{y}, 
      \PPKp{x}{y}, \f{x}{y}, 0, 1}\\
  \end{array}
\right.
$$

Note that the presentation \call{A_{DEO}} is not convergent, and thus
we cannot apply results on basic narrowing as is. To this end we
introduce a rewriting system $\call{R}_{DEO}$ which is convergent and obtained
by Knuth-Bendix~\cite{KB} completion on \call{A_{DEO}}, and such that
two terms have the same normal form for $\call{R}_{DEO}$ iff they are equal
modulo \call{H_{DEO}}.

\begin{lemma}{\label{lemma:00:convergent}}
  \call{H_{DEO}} is generated by the convergent rewriting system:
  $$
  \call{R}_{DEO}=\left\lbrace
    \begin{array}[c]{l}
      \Ver{x}{\Sig{x}{\SK{y}}}{\PK{y}} \to 1\\
      \Ver{x}{\Sig{x}{\SSKp{y_1}{y_2}}}{\PPKp {y_1}{y_2}} \to 1\\
      \Ver{\f{\PK{y}}{\Sig{x}{\SK{y}}}}{\Sig{x}{\SK{y}} }  {\PPKp  {\PK{y}}{\Sig{x}{\SK{y}} }} \to 1 \\
      \Sig{\f{\PK{y}}{\Sig{x}{\SK{y}}}} { \SSKp {\PK{y}}{\Sig{x}{\SK{y}} }  }\to \Sig{x}{\SK{y}} 
    \end{array}
  \right.
  $$
\end{lemma}

\section{Decidability of unifiability\label{unifdecidability}}
It can easily be shown, using the criterion of termination of basic
narrowing on the right-hand side of rules of $\call{R}_{DSKS}$ (resp. $\call{R}_{DEO}$), that basic
narrowing terminates when applied with the rules of $\call{R}_{DSKS}$ (resp. the rules of $\call{R}_{DEO}$).  The
main result of~\cite{Hullot} then implies the following proposition,
when applying basic narrowing with $\call{R}_{DSKS}$ (resp. $\call{R}_{DEO}$) non-deterministically on
the two sides of an equation modulo $\call{R}_{DSKS}$ (resp. $\call{R}_{DEO}$) and terminates with
unification modulo the empty theory.

\begin{proposition}{\label{theo:term}}
  Basic narrowing is a sound, complete and terminating procedure for
  finding a complete set of most general $\call{H_{DSKS}}$-unifiers (resp. $\call{H_{DEO}}$ unifiers).
\end{proposition}

One can actually be more precise, and we will need the following
direct consequence of Hullot's unification procedure, that states that
applying basic narrowing permits one to ``guess'' partially the normal
form of a term $t$.

\begin{lemma}{\label{lemma:00:narrow}}
  Let $t$ be any term and $\sigma$ be a normalised substitution.
  There exists a term $t'$ and a substitution $\sigma'$ in normal form
  such that $t\bnar^* t'$ and $t'\sigma'=\norm{t\sigma}$ where $\bnar$ represent a basic narrowing relation modulo \call{H_{DSKS}} (resp.  modulo \call{H_{DEO}}).
\end{lemma}

While this presentation by a convergent rewrite system ensures the
decidability of unification modulo \call{H_{DSKS}} (resp. modulo \call{H_{DEO}}), we prove bellow  that the unifiability problem, as well as the partial
guess of a normal form, is in fact in NPTIME.

\paragraph{\bf{Complexity of unification}}
\begin{theorem}{\label{theo:lenderiv}}
  Let $t$ be a term and $(D)$ be a basic narrowing derivation modulo
  $\call{H_{DSKS}}$ (resp. modulo $\call{H_{DEO}}$) starting from $t$.  Then, the length of $(D)$ in
  bounded by $\|t\|.$
\end{theorem}

\begin{proof}
 Let us prove the theorem for the basic narrowing derivations modulo $\call{H_{DSKS}}$. Let $t$ be a term and
  $D$ be a basic narrowing derivation starting from $t$, $D: t=t_0
  \leadsto_{b.n} t_1\leadsto_{b.n} \ldots \leadsto_{b.n} t_n.$
  $\mathcal{R}_{DSKS}$ is convergent and any basic narrowing derivation
  starting from the right members of the rules of $\mathcal{R}_{DSKS}$
  terminates, then, by \cite{Hullot}, $(D)$ terminates.  Let us prove
  that $\|D\| \leq \|t\|.$ Let $P_0=\PP{t_0}$ be the number of
  distinct subterms of $t_0$ where we can apply the basic narrowing.
  We note that if the basic narrowing can be applied on a term $s$ at
  a position $p$ and if there exists another subterm of $s$ at
  position $q$ such that $t_{|p}=t_{|q}$, we apply the basic narrowing
  at the positions $p$ and $q$ at the same time.  Let $t_i
  \leadsto_{b.n} t_{i+1}$ be a step in $(D)$ and let $l_i\to
  r_i\in\mathcal{R}_{DSKS}$ be the applied rule.  For any $l\to r \in
  \call{R}_{DSKS}$, $r$ is not narrowable.  By the fact that $r_i$ is not
  narrowable and by the definition of basic narrowing \cite{Hullot},
  we have $P_{i+1} < P_i.$ We deduce that $\|D\| \leq P_0$, but
  $P_0\leq \|t_0\|$ then $\|D\| \leq \|t\|.$

 The case of derivations modulo $\call{H_{DEO}}$ is analogoes. 
\end{proof}

\begin{corollary}{\label{coro:cor1}}
  The \call{H_{DSKS}}-unifiability (resp. \call{H_{DEO}}-unifiability) can be decided in NPTIME.
\end{corollary}

\begin{proof}
Let us prove the corollary for \call{H_{DSKS}}-unifiability.
  Let $P$ and $Q$ be two terms. $\mathcal{R}_{DSKS}$ is convergent and any
  basic narrowing derivation starting from the right members of the
  rules of $\mathcal{R}_{DSKS}$ terminates then, there exists an
  $\call{H}_{DSKS}$-unification algorithm (proposition
  \ref{theo:term}).  Let us prove that this algorithm runs in NPtime.
  Suppose $M=H(P,Q)$, ($H$ is a new function symbol representing the
  cartesian product), and $m=\|M\|=\|P\|+\|Q\|+1$.  For any basic
  narrowing derivation $D$ starting from M, we have $\|D\| \leq m$
  (Theorem \ref{theo:lenderiv}).  Suppose that our algorithm always
  explore the right branch, then, starting from any term $M$, the
  algorithm will be perform at most $\|M\|$ steps before halting.
  Then, we have the corollary.

 By the same reasoning, we prove that \call{H_{DEO}}-unifiability can be decided in NPTIME.
\end{proof}


\section{Saturation}
\label{sec:saturation}

\subsection{Construction}
Let $\call{H}$ be an equational theory presented by a convergent rewrite system \call{R}.
The saturation of the set of deduction rules \call{L} defined modulo the equational theory \call{H} is the output of the application of
the saturation rules of Figure~\ref{fig:saturation} starting with
$\call{L}'=\call{L}$ until any added rule is subsumed by
a rule already present in $\call{L}'$.

\begin{figure*}[htbp]
  $$
  \begin{array}{lc}
    \mathsf{Subsumption}:&
    \hspace*{1em}%
    \vcenter{
      \infer
      [l_1\subseteq l_2]
      {
        \call{L}'\gets \call{L}'\setminus\set{l_2\ded r}
      }%
      {
        l_1\ded r\in\call{L}'~~~~l_2\ded r\in\call{L}'
      }
    }\\[1em]
    \mathsf{Closure}:&
      \infer
      [\begin{array}{c}
        t\notin \Variables\\
        \sigma=mgu_\emptyset(r_1,t)\\
      \end{array}]
      {
        \call{L}'\gets\call{L}'\cup\set{(l_1,l_2\ded r_2)\sigma}
      }
      {
        l_1\ded r_1 \in \mathcal{L}',~~~~ (t,l_2) \ded r_2\in \mathcal{L}'
      }
    \\[1em]
    \mathsf{Narrow}:&
    \hspace*{-2.5em}
    \vcenter{
      \infer
      {
        \call{L}'\gets\call{L}'\cup\set{l'\ded r'}
      }
      {
        l\ded r\in\call{L}'~~~~(l,r)\bnar (l',r')
      }
    }
  \end{array}
  $$
  \vspace*{-1em}

  \caption{\label{fig:saturation} System of saturation rules.}
  \vspace*{-2em}  
\end{figure*}

The application of the saturation rules on $\call{L_{DSKS}}$ (resp. on $\call{L_{DEO}}$) terminates, and yields the
following sets of rules:\\
$\call{L_{DSKS}}'=\call{L_{DSKS}} \cup x,\SK{y}\ded {\Sig{x}{\SK{y}}}\cup x, \SKp{\PK{y}}{\Sig{x}{\SK{y}}}\ded \Sig{x}{\SK{y}}$
and \\ 
$\call{L_{DEO}}'=\call{L_{DEO}} \cup \f{\PK{y}}{\Sig{x}{\SK{y}}},\SSKp{\PK{y}}{\Sig{x}{\SK{y}}}  \ded \Sig{x}{\SK{y}}$

We define four new \emph{extended} intruder systems: $\IIDSKS=
\intrus{\call{G_{DSKS}}}{\call{L_{DSKS}}'}{\call{H}_{DSKS}}$,
$\mathcal{I}_{\emptyset 1}=\intrus{\call{G_{DSKS}}}{\call{L_{DSKS}}'}{\emptyset}$, $\IIDEO=
\intrus{\call{G_{DEO}}}{\call{L_{DEO}}'}{\call{H}_{DEO}}$ and
$\mathcal{I}_{\emptyset 2}=\intrus{\call{G_{DEO}}}{\call{L_{DEO}}'}{\emptyset}$.
These intruder systems do not satisfy the requirements that the
left-hand side of deduction rules have to be variables.  The deduction
relation, the derivations and the set of reachable terms are defined
as usual from ground instances of deduction rules.

\subsection{Properties of a saturated system}
In the rest of this paper, we suppose \call{H}, \call{R}, \call{L}, \call{L}',  $\I=\intrus{\call{G}}{\call{L}}{\call{H}}$, 
$\II=\intrus{\call{G}}{\call{L}'}{\call{H}}$ and $\mathcal{I}_\emptyset=\intrus{\call{G}}{\call{L}'}{\emptyset}$ to be either respectively  \call{H_{DSKS}}, \call{R_{DSKS}}, \call{L_{DSKS}}, \call{L_{DSKS}'}, \IDSKS, \IIDSKS    
or respectively  \call{H_{DEO}}, \call{R_{DEO}}, \call{L_{DEO}}, \call{L_{DEO}'}, \IDEO, \IIDEO.

Let us first prove that the deduction system obtained after saturation
gives exactly the same deductive power to an intruder.

\begin{lemma}{\label{lemma:00:saturation1}}
  For any set of normal ground terms $E$ and any normal ground term
  $t$ we have: $E\ded^*_{\I} t$ if and only if $E\ded^*_{\II}
  t$.
\end{lemma}

\begin{proof}
  First, let us assume that $t\in\rhclos{E}{\I}$, that is, there exists a $\I$-derivation $(D)$ starting from $E$ of goal $t$, and let us
  prove that there exists a $\II$-derivation starting from  $E$ of goal $t$.
  If there exists a step in the
  derivation $D$ which uses a rule $l\ded r \in \call{L}$ but not in
  $\call{L}'$, then, by construction of $\call{L}'$, there
  exists another rule $l_1\ded r$ in $\call{L}'$ such that
  $l_1\subseteq l$ and thus that can be applied instead of $l\ded r$.
  We conclude that
  $E\ded^*_{\II} t$.
  
  For the reciprocal, let us assume that there exists a $\II$-derivation
  starting from $E$ of goal $t$, and
  let us prove that there exists a $\I$-derivation  starting from $E$
  of goal $t$.  We begin by defining an
  arbitrary order on the rules of $\call{L}$, and we extend this
  order to the rules of $\call{L}'\setminus\call{L}$ as follows:
  the rules of $\call{L}$ are smaller than the rules of
  $\call{L}'\setminus\call{L}$ and the rules of
  $\call{L}'\setminus\call{L}$ are ordered according to the order of their
  construction during the saturation.  Let $\M{D}$ be the
  multiset of deduction rules applied in $D$.  
  Let $\Omega(E,t)=\set{D\mid D:E\ded^*t}$. Since $t\in\rhclos{E}{\II}$,
  $\Omega(E,t)\not=\emptyset$.
  Let $D$ be a derivation in $\Omega(E,t)$ having the
  minimal $\M{D}$, and let us prove that $D$ does not use rules in
  $\call{L}'\setminus\call{L}$. By contradiction, suppose that $D$ uses a
  rule $l\ded r \in\call{L}'\setminus\call{L}$. 
  Since $l\ded r\notin\call{L}$, it has been constructed according to the rules of saturation. Let us review the possible cases:
  \begin{itemize}
  \item If $l\ded r$ has been constructed by the third rule of saturation,
    there exists a rule $l_1 \ded r_1\in\call{L}'$ such
    that $(l_1,r_1)\leadsto^*_{b.n}(l,r)$.  By definition of deductions, $l_1\ded r_1$ can be
    applied instead of $l\ded r$. Let $(D')$ be the derivation where
    $l'\ded r'$ replaces $l\ded r$, $(D')$ is in $\Omega(E,t)$.  Since $l_1\ded r_1$ has an order smaller than the
    order of $l\ded r$, we have  $\M{D'} < \M{D}$, which contradicts the
    minimality of $\M{D}$.
  \item If $l\ded r$ has been constructed by the second rule of
    saturation, there exists two rules $l_1\ded
    r_1$ and $s,l_2\ded r_2$ in $\call{L}'$ such that
    $\mu=mgu(r_1,s)$, $s\notin\Variables$, $l=\norm{(l_1,l_2)\mu}$ and
    $r=\norm{r_2\mu}$.  suppose that $l\ded r$ is applied on the set of terms $F$,
    $F\ded_{l\ded r} F,g$. Since  $\norm{l\sigma}\subseteq F$ and $\norm{r\sigma} = g$ for a substitution $\sigma$, we have $\norm{l_1\mu\sigma}\subseteq F$ and $\norm{(s,l_2)\mu\sigma}\subseteq F\cup\norm{r_1\mu\sigma}$, this implies that 
    $F\ded_{l_1\to r_1}F,\norm{r_1\mu\sigma}\ded_{l_2,s\to r_2} F,\norm{r_1\mu\sigma},g$.
    Let $(D')$ be the derivation where
    $l_1\ded r_1$ and $s,l_2\ded r_2$ replace $l\ded r$. $(D')$ is in $\Omega(E,t)$.   Since $l_1\ded r_1$ and $s,l_2\ded r_2$ have an order smaller than the order 
    of $l\ded r$, we have  $\M{D'} < \M{D}$ which
    contradicts the minimality of $\M{D}$.
  \end{itemize}
  We conclude that $(D)$ does not use rules in
  $\call{L}'\setminus\call{L}$, then, we have the
  reciprocal of the lemma.
\end{proof}

Moreover, we can prove that when considering only deductions on terms
in normal form and yielding terms in normal form, it is sufficient to
consider derivations modulo the empty theory
(Corollary~\ref{coro:cor2}).

\begin{lemma}{\label{lemma:00:lem1}}
  Let $E$ (resp. $t$) be a set of terms (resp. a term) in normal form.
  We have: $E\ded_{\II} E,t$ if and only if
  $E\ded_{\mathcal{I}_\emptyset} E,t$.
\end{lemma}
\begin{proof}
 Assume first $E\ded_{\II} E,t$. There exists a rule $l\ded r \in
  \call{L}'$ and a substitution $\sigma$ in normal form such
  that $\norm{l\sigma}\subseteq E$ and $t=\norm{r\sigma}$.  By
  Lemma~\ref{lemma:00:narrow}, there exists a set of terms $l'$, $r'$,
  and a substitution $\sigma'$ in normal form such that $l\bnar^* l'$,
  $r\bnar^* r'$, and $\norm{l\sigma}=l'\sigma'$, and
  $\norm{r\sigma}=r'\sigma'$.  By the saturation, we have added at
  some point $l'\ded r'$ to $\mathcal{L}'$. Either this rule is
  present in the final \call{L}' and can be applied, or it is
  subsumed by a rule that can be applied on $E$. The converse is left
  to the reader.\end{proof}

\begin{corollary}{\label{coro:cor2}}
  Let $E$ (resp. $t$) be a set of terms (resp. a term) in normal form.
  We have: $E\ded^*_{\II} E,t$ if and only if
  $E\ded^*_{\mathcal{I}_\emptyset} E,t$.
\end{corollary}

Next lemma states that if a term in the left-hand side of a deduction
rule of the saturated system is not a variable, then we can assume it
is not the result of another saturated deduction rule.

\begin{lemma}{\label{lemma:00:saturation2}}
  Let $E$ (resp. $t$) be a set of terms (resp. a term) in normal form.
  If $t\in\rhclos{E}{\call{I}_\emptyset}$, then there exists a
  $\call{I}_\emptyset$-derivation starting from $E$ of goal $t$ such that:\\
  for all $\call{I}_\emptyset$ rules $l\ded r$ applied with substitution $\sigma$,
  for all $s\in l\setminus\Variables$, we have $s\sigma\subseteq E$.
\end{lemma}
\begin{proof}
 Let us prove by induction
 on the length $n$ of a derivation $D$ starting from $E$ of goal $t$ that 
 either $D$ satisfies the property or their exists
 another $\call{I}_\emptyset$-derivation $D'$ of length smaller  to $n$  starting from $E$ of goal $t$ 
 which satisfies
 the property.\\
 The case $n=1$ is obvious.\\
 Suppose that the lemma is true for derivations of length $\leq n$  and
 let us prove it for  derivations $D$ of length $n+1$.\\
 $D: E=E_0\ded^{i-1} E_{i-1}\ded E_{i-1},t_i \ded\ldots\ded E_{n} \ded
 E_{n},t$.  Suppose that $D$ does not satisfy the property, there exists a step $i$ in $D$ 
 where the  rule $l\ded r$ is applied with  the substitution  $\sigma$, and there exists  $s\in l\setminus\Variables$ such that   $s\sigma\notin E$.  
 Since $s\sigma\notin E$, it has been constructed at some step $j<i$. We have:\\ 
 $D:~ E \ded^{j-1} E_{j-1} \ded E_{j-1},s\sigma\ded \ldots \ded E_i\ded
 E_i,t_i\ded\ldots \ded E_{n} \ded E_{n},t$.  
 Let $l_j\ded r_j\in\mathcal{L}'$  be the rule applied, with the substitution
 $\tau$, to construct $s\sigma$. Since $r_j\tau=s\sigma$, $r_j$ and $s$ are unifiable with
 $\mu=mgu(r_j,s)$. Then the rule $(l_j,l\setminus s\ded r)\mu$ has been constructed.
 Since $\mu$ is a substitution  most general  than $\sigma$, it can be applied on $E_{i}$
 to yield $t_i$.  
 This implies that we can reduce $D$ to:\\
 $D':~ E \ded^j E_j \ded E_{j+2} \ded \ldots \ded E_i\ded E_i,t_i\ded\ldots
 \ded E_{n} \ded E_{n},t$ where the construction of $s\sigma$ is
 spell and the applied rule at the step $i$ is $(l_j,l\setminus s)\mu \ded
 r\mu$.  We note that $\|D'\| < \|D\|,$ then $\|D'\|\leq n.$ By
 induction, either $D'$ satisfies the property or there exists another $\call{I}_\emptyset$-derivation $D''$ starting from $E$ of goal $t$  which satisfies the property. Then, we have the lemma
for derivations of  length n+1 and this concludes the proof.
\end{proof}

\section{Decidability of reachability}\label{sec:decision}

The main result of this paper is the following theorem.

\begin{theo}
  The \IDSKS-\textsf{Reachability} (resp. \IDEO-\textsf{Reachability}) problem is decidable.
\end{theo}

The rest of this paper is devoted to the presentation of an algorithm
for solving \IDSKS-\textsf{Reachability} (resp. \IDEO-\textsf{Reachability}) problems and to a proof scheme of its completeness, correctness and  termination. This decision procedure comprises three different
steps.

Let \call{C} be an \call{I}-constraint system.
\subsection{First step: guess of a normal form}
\paragraph{Step 1.} 
Apply non-deterministically basic narrowing steps on all subterms of
$\call{C}$. Let $\call{C}_0=\set{(E_i^0 \rhd{}
  v_i^0)_{i\in\set{1,\ldots,n}} ,\call{S}^0}$ be the resulting
constraint system.

\paragraph{Remark.} Let $\sigma$ be a solution of the original
constraint system, with $\sigma$ in normal form. This first step will
non-deterministically transform each $t\in\Sub{\call{C}}$ into a term
$t'$ such that, according to Lemma~\ref{lemma:00:narrow} we will have
$\norm{t\sigma}=t'\sigma'$.

\subsection{Second step:  resolution of unification problems}

\paragraph{Step 2.}
Solve the unification system $\call{S}^0$ modulo the empty theory, and
apply the obtained unifier on the deduction constraints to obtain a
constraint system $\call{C}'=\set{(E_i'\rhd{}
  t_i')_{i\in\set{1,\ldots,n}}}$

\paragraph{Remarks.} We prove below that if there exists a solution to
the original constraint system, then there exists a solution of
$\call{C}'$ for the extended intruder system $\call{I}_\emptyset$.
\call{C}' itself is not a constraint system, but an \emph{extended}
constraint system.

\begin{lemma}{\label{lemma:00:lem2}}
  If $\sigma$ is a substitution in normal form such that
  $\sigma\models_{\I}\call{C}$, there exists a $\call{C}'$ at
  Step~2 and a substitution $\sigma'$ in normal form such that
  $\call{C}\bnar^*\call{C}'$ and
  $\sigma'\models_{\mathcal{I}_\emptyset}\call{C}'$.
\end{lemma}
\begin{proof} By definition $\sigma\models_{\I}\call{C}$ implies that for all
  $i\in\set{1,\ldots,n}$ we have $\sigma\models_{\I}
  (E_i\rhd{}t_i)$. Thus there exists by
  Lemma~\ref{lemma:00:saturation1} an $\II$-derivation starting
  from $\norm{E_i\sigma}$ to $\norm{t_i\sigma}$.  Since $\sigma$ is in
  normal form, by lemma \ref{lemma:00:narrow}, there exists $E'_i$,
  $t'_i$ and $\sigma'$ in normal form such that $E_i\bnar^* E'_i$,
  $t_i\bnar^* t'_i$, $\norm{E_i\sigma}=E'_i\sigma'$ and
  $\norm{t_i\sigma}=t'_i\sigma'$ for all $i\in\set{1,\ldots,n}$.  By
  Lemma \ref{lemma:00:lem1}
  $\norm{E_i\sigma}\to^*_{\II}\norm{t_i\sigma}$ then implies
  $\norm{E_i\sigma}\to^*_{\mathcal{I}_\emptyset}\norm{t_i\sigma}$
  Since $\norm{E_i\sigma}=E'_i\sigma'$ and
  $\norm{t_i\sigma}=t'_i\sigma'$ then,
  $\sigma'\models_{\mathcal{I}_\emptyset}(E'_i\rhd{}t'_i)$ for all
  $i\in\set{1,\ldots,n}$ and thus we have the lemma. \end{proof}

\subsection{Third step: Transformation in solved form}

\begin{figure*}[htbp]
  $$
  \begin{array}{lc}
    \mathsf{Apply}:&\\
    \multicolumn{2}{c}{%
      \infer
      [\begin{array}{c}
        l_x,l_1,\ldots,l_n\ded r\in\call{L}'\text{ and }
        l_x\subseteq\Variables, t\notin\Variables\\
        e_1,\ldots,e_n\in E\text{ and }
        \sigma=mgu(\set{(e_i\unif{}l_i)_i,r\unif{}t})\\
      \end{array}]
      {
        (\call{C}_\alpha, (E\rhd{} y)_{y\in l_x},  \call{C}_\beta)\sigma
      }
      {
        \call{C}_\alpha,
        E\rhd{}t,
        \call{C}_\beta
      }
    }\\[1em]
    \mathsf{Unif}:&
    \hspace*{-2.5em}
    \vcenter{
      \infer
      [
      \begin{array}{c}
        u,t\notin\Variables\\
        u\in{}E,~ \sigma= mgu{}(u,t)
      \end{array}
      ]%
      {
        (\call{C}_\alpha,\call{C}_\beta)\sigma
      }
      {
        \call{C}_\alpha,E\rhd{}t,\call{C}_\beta
      }
    }
  \end{array}
  $$
  \vspace*{-1em}

  \caption{\label{fig:resolution} System of transformation rules.}
  \vspace*{-2em}
\end{figure*}

 \paragraph{Step 3.} 
 To simplify the constraint system, we apply the transformation rules
 of Figure~\ref{fig:resolution}. Our goal is to transform \call{C}'
 into a constraint system such that the right-hand sides of deduction
 constraints (the $t_i$) are all variables. When this is the case, we
 say that the constraint system is in \emph{solved form}. It is
 routine to check that a constraint system in solved form is
 satisfiable.




%


\begin{lemma}{\label{lemma:00:eliminationvariables}}
  Let $\call{C}=\set{\call{C}_\alpha,E\rhd{}t,\call{C}_\beta}$ be such
  that $\call{C}_\alpha$ is in solved form.  Then, for all substitution
  $\sigma$, $\sigma\models\call{C}$ if and only if $\sigma\models
  \set{\call{C}_\alpha,(E\setminus \Variables)\rhd{}t,\call{C}_\beta}$ .
\end{lemma}
\begin{proof} It suffices to prove that if $x\in E\cap\Variables$ and $\sigma$ is
  a substitution such that $\sigma\models\call{C}$, then we have
  $\sigma\models\set{\call{C}_\alpha,(E\setminus
    \set{x})\rhd{}t,\call{C}_\beta}$.  Given $x\in E$ there exists a
  set of terms $E_x\subseteq E$ such that
  $E_x\rhd{}x\in\call{C}_\alpha$.  Since $\sigma\models\call{C}$ we
  have $\sigma\models E_x\rhd{}x$, and by the fact that $E_x\subseteq
  E\setminus\set{x}$ we have $\sigma\models E\setminus\set{x}\rhd{}x$.
  Since we also have $\sigma\models (E\rhd{}t) $ this implies
  $\sigma\models E\setminus\set{x}\rhd{}t$.  The reciprocal is obvious
  since $E\setminus\set{x}\subseteq E$. \end{proof}

It also can be proved that the lazy constraint solving procedure
terminates. This lemma also helps us to prove the completeness of lazy
constraint solving (stated in Lemma~\ref{lemma:00:lem3}).

\begin{lemma}{\label{lemma:00:terminaison1}}(DSKS-termination.)
  Let $\call{C}$ be an $\mathcal{I}_{DSKS}$-constraint~ system.  The application of
  transformation rules of the \emph{algorithm} using  \call{L_{DSKS}'} rules terminates.
\end{lemma}

\begin{proof}
Let $\nbv{\call{C}}=|\Var{\call{C}}|$ be the number of variables in
  \call{C}, and $\call{M}(\call{C})$ denote the multiset of the
  right-hand side of deduction constraints in \call{C}. Let us prove
  that after any application of a transformation rule on a constraint
  system $\call{C}=(\call{C}_\alpha,E\rhd t)$ (where $\call{C}_\alpha$
is in solved form), either \nbv{\call{C}} decreases strictly, or the
identity substitution is applied on \call{C} during the transformation
and $\call{M}(\call{C})$ strictly decreases.

The first point will ensure that after some point in a sequence of
transformations the number of variables will be stable, and thus from
this point on $\call{M}(\call{C})$ will strictly decrease. The fact
that no more unification will be applied and that the extension of the
subterm ordering on multisets is well-founded will then imply that
there is only a finite sequence of different constraint systems, and
thereby the termination of the constraint solving algorithm.

This fact is obvious if the \textsf{Unif} rule is applied, since it
amounts to the unification of two subterms of \call{C}. It is then
well-known that if the two subterms are not syntactically equal, the
number of variables in their most general unifier is strictly less
than the union of their variables, which is included in
\Var{\call{C}}. If they are syntactically equal, then no substitution
is applied, and thus denoting \call{C}' the result of the
transformation, we have
$\call{M}(\call{C})=\call{M}(\call{C}')\cup\set{t}$, and thus
$\call{M}(\call{C'})<\call{M}(\call{C})$.

Let us now consider the case of the \textsf{Apply} rule, and let
\call{C}' be the obtained constraint system. If the underlying
intruder deduction rule is in \call{L_{DSKS}}, the fact that $t$ is
not a variable implies that the variables of the right-hand side of
the rule will be instantiated by the strict maximal subterms
$t_1,\ldots,t_k$ of $t$. We will thus have:
$$
\call{M}(\call{C}')=\call{M}(\call{C})\cup\set{t_1,\ldots,t_n}\setminus
\set{t}
$$
and thus $\call{M}(\call{C}')<\call{M}(\call{C})$.

It now suffices to prove the Lemma for the two rules in
$\call{L_{DSKS}'}\setminus\call{L_{DSKS}}$:
\begin{description}
\item[rule $x,\SK{y}\ded \Sig{x}{\SK{y}}$:] The substitution $\sigma$
  applied is the most general unifier of the unification system
  \set{\Sig{x}{\SK{y}}\unif{}t,\SK{y}\unif{}u} for some $u\in E$.
  Since this is syntactic unification and since we can assume neither
  $u$ (by Lemma~\ref{lemma:00:eliminationvariables}) nor $t$ (by
  definition of the \textsf{Apply} rule) are variables, we must have
  $u=\SK{u'}$ and $t=\Sig{t_1}{t_2}$. The second equation thus yields
  $y=u'$, with $u'\in\Sub{\call{C}}$. Replacing in the first equation,
  $\sigma$ is the most general unifier of the equation
  $\Sig{x}{\SK{u'}}\unif{}\Sig{t_1}{t_2}$, which reduces into the set
  of equations \set{x\unif{}t_1,\SK{u'}\unif{}t_2}. The first equation
  implies that $x$ is instantiated by a strict subterm $t_1$ of $t$.
  If the second equation is trivial we have
  $\call{M}(\call{C}')=\call{M}(\call{C})\cup
  \set{t_1}\setminus\set{t}$, and thus
  $\call{M}(\call{C}')<\call{M}(\call{C})$. Otherwise, since
  $\Var{\SK{u'}}\cup\Var{t_2}\subseteq\Var{\call{C}}$ we have
  $\nbv{\call{C}'}<\nbv{\call{C}}$.
\item[rule $x,\SKp{\PK{y}}{\Sig{x}{\SK{y}}}\ded \Sig{x}{\SK{y}}$:] The
  substitution $\sigma$ applied is the most general unifier of the
  unification system
  \set{\Sig{x}{\SK{y}}\unif{}t,\SKp{\PK{y}}{\Sig{x}{\SK{y}}}\unif{}u}
  for some $u\in E$.  Since this is syntactic unification and since we
  can assume neither $u$ (by
  Lemma~\ref{lemma:00:eliminationvariables}) nor $t$ (by definition of
  the \textsf{Apply} rule) are variables, we must have
  $u=\SKp{u'_1}{u_2'}$ and $t=\Sig{t_1}{t_2}$. If $\sigma$ is the
  identity on \call{C}, we are done, since in this case we have
  $\call{M}(\call{C}')=\call{M}(\call{C})\cup\set{t_1}\setminus\set{t}$
  and thus $\call{M}(\call{C}')<\call{M}(\call{C})$. Otherwise let us
  examine how the unification system is solved. It is first
  transformed into:
  $$
  \set{x\unif{}t_1, \SK{y}\unif{}t_2,\PK{y}\unif{}u_1',\Sig{x}{\SK{y}}
    \unif {u_2'}}
  $$
  Resolving the first equation yields (note that $x\notin\Var{\call{C}}$):
  $$
  \set{\SK{y}\unif{}t_2,\PK{y}\unif{}u_1',\Sig{t_1}{\SK{y}} \unif
    {u_2'}}
  $$
  Let us consider two cases, depending on whether both $u_1'$ and $t_2$ are variables:
  \begin{itemize}
  \item If they are both variables, then solving the first equation
    removes $t_2$ from \Var{\call{C}} but adds a variable $y$. The
    second equation will also remove $u_1'$, but since the variable
    $y$ is already present, it will not add another variable. Since
    \PK{y} and \SK{y} are not unifiable, we note that we must have
    $t_2\neq u_1'$, and thus we have removed two variables and added
    one by solving the two first equations. The remaining equation
    contains only variables of the ``intermediate'' constraint system,
    and thus will not add any new variable. In conclusion, in this
    case, the number of variables of \call{C} decreases by at least 1.
  \item If say $t_2$ is not a variable, and thus $t_2=\SK{t_2'}$, with
    $t_2'\in\Sub{\call{C}}$. Resolving the first equation and
    injecting the solution in the remaining equations yields the
    unification system:
    $$
    \set{\PK{t_2'}\unif{}u_1',\Sig{t_1}{\SK{t_2'}} \unif {u_2'}}
    $$
    Note that up to this point the substitution $\sigma$ that we built
    does not affect any variable of $\call{C}$. If this remaining
    unification system is trivial, then the substitution applied on
    \call{C} is the identity, we are done (see above). Otherwise,
    since all the variables in this system are in \Var{\call{C}}, it
    strictly reduces \nbv{\call{C}}. This terminates the proof of this
    case.
  \end{itemize}
  Thus, if this rule is applied, either no substitution is applied on
  \call{C} and $\call{M}(\call{C})$ strictly decreases, or the number
  of variables in the resulting constraint system \call{C}' is
  strictly smaller than the number of variables in \call{C}.
\end{description}

\end{proof}

\begin{lemma}{\label{lemma:00:terminaison2}}(DEO-termination.)
  Let $\call{C}$ be an $\mathcal{I}_{DEO}$-constraint~ system.  The application of
  transformation rules of the \emph{algorithm} using $\call{L_{DEO}}'$ rules terminates.
\end{lemma}
\begin{proof}
Let  $\call{C}=\set{(E_i \rhd{} t_i)_{i\in\set{1,\ldots,n}}}$ 
be an $\mathcal{I}_{DEO}$-constraint system not in solved form
and let the complexity of $\call{C}$  be a couple  ordered lexicographically with  the following components:
\begin{itemize}
\item
$\nbv{\call{C}},$  the number of distinct variables in $\call{C}$,
\item 
$\call{M}(\call{C})$ the multiset of the right-hand side of deduction constraints in \call{C}.
\end{itemize}

We have to show that each rule reduces the complexity.
The fact is obvious if the $Unif$ rule is applied, since it amounts to the unification of two subterms of \call{C}.
If is then well-known that if two subterms are not syntactically equal, then the number of variables in their most general unifier is strictly less than the union of their variables, which is included in $\Var{\call{C}}$. If their are syntactically equal, then no substitution is applied, and thus denoting \call{C}' the result of transformation, we have
$\call{M}(\call{C}') < \call{M}(\call{C})$.

Let us now consider the case of $Apply$ rule, and let \call{C}' be the obtained constraint system. If the underlying intruder deduction rule is in $\mathcal{L}_{DEO}$, the fact  that $t$ is not a variable implies that the right-hand side of the rule will be instantiated by the strict maximal subterms $t_1,\ldots, t_k$ of t. we will thus have 
$\call{M}(\call{C}')=\call{M}(\call{C})\cup\set{t_1,\ldots,t_k}\setminus \set{t}$ and thus $\call{M}(\call{C}')<\call{M}(\call{C})$.

It is now suffices to prove the Lemma for the rule in $\mathcal{L_{DEO}}'\setminus\mathcal{L_{DEO}}$:

the applied rule is:  $\f{\PK{y}}{\Sig{x}{\SK{y}}},\SSKp{\PK{y}}{\Sig{x}{\SK{y}} }  \ded \Sig{x}{\SK{y}}$.
The substitution $\sigma$ is the most general unifier of the unification system 
$\set{t\unif{}\Sig{x}{\SK{y}}, e_1\unif{} \f{\PK{y}}{\Sig{x}{\SK{y}}}, e_2\unif{} \SSKp{\PK{y}}{\Sig{x}{\SK{y}}}}$ for some $e_1,~ e_2\in E.$ since it is syntactic unification and since we can assume neither $e_1$, neither $e_2$ (by Lemma \ref{lemma:00:eliminationvariables}) nor $t$ (by definition of the Apply rule) are variables, we must have 
$t=\Sig{t_1}{t_2}, ~ e_1=\f{v_1}{v_2},$ and $e_2=\SSKp{v_3}{v_4}$.
\begin{itemize}
\item
If $t_2\in\Variables$, we have  $\sigma(x)=t_1, ~  \sigma(t_2)=\SK{y}$ and the unification system is then transformed into:

$\set{v_1\unif{}\PK{y}, v_2\unif{} \Sig{t_1}{\SK{y}}, v_3\unif{}\PK{y}, v_4\unif{}\Sig{t_1}{\SK{y}}}$.

By the fact that $t_3$ is replaced by $y$, $x,y\notin\Var{\call{C}}$, and  the number of variables in $\sigma$  is strictly less than the union of variables of the unification system, we deduce that $\nbv{\call{C}'}<\nbv{\call{C}}$. 

\item
If $t_2\notin\Variables$ then $t_2=\SK{t_3}$.
We have $\sigma(x)=t_1,  ~ \sigma{y}= t_3$ and the unification system
is then transformed into:

$\set{v_1\unif{}\PK{t_3}, v_2\unif{}\Sig{t_1}{\SK{t_3}}, v_3\unif{}\PK{t_3}, v_4\unif{}\Sig{t_1}{\SK{t_3}}}$.

If the unification system is obvious, that is $\sigma$ is the identity substitution, we have $\call{C}'=\call{C}\setminus (E\rhd{}t)$, and then $\call{M}(\call{C}')=\call{M}(\call{C})\setminus{t}$ which implies that $\call{M}(\call{C}')<\call{M}(\call{C})$.
Else, we have $\nbv{\call{C}'}<\nbv{\call{C}}$. 
\end{itemize}
This concludes the proof.
\end{proof}

\begin{lemma}{\label{lemma:00:lem3}}
  If $\call{C}'$ is satisfied by a substitution $\sigma'$, it can be
  transformed into a system in solved form by the rules of
  Figure~\ref{fig:resolution}.
\end{lemma}

\begin{proof}
  Let $\call{C}$ be a deterministic constraint system not in solved
  form and let $i$ be the smallest integer such that
  $t_i\notin\Variables$, then
  $\call{C}=\set{\call{C}_\alpha,E_i\rhd{}t_i,\call{C}_\beta}$ where
  $\call{C}_\alpha$ is in solved form.  Let $\sigma$ be a substitution
  such that $\sigma\models_{\mathcal{I}_{\emptyset}}\call{C}$, and let
  us prove that $\call{C}$ can be reduced to another satisfiable constraint system
  $\call{C}'$ by applying the transformation rules given in the
  algorithm.  $\sigma\models_{\mathcal{I}_{\emptyset}}\call{C}$, then
  $\sigma\models_{\mathcal{I}_{\emptyset}}\set{\call{C}_\alpha,E_i\setminus\Variables\rhd{}t_i,\call{C}_\beta}$
  (Lemma \ref{lemma:00:eliminationvariables}) and then,
  $(E_i\setminus\Variables)\sigma\to^*_{\mathcal{I}_{\emptyset}}t_i\sigma$.
  We have two cases:
  \begin{itemize}
  \item If $t_i\sigma\in (E_i\setminus\Variables)\sigma$,  there
    exists a term $u\in (E_i\setminus\Variables)$ such that
    $u\sigma=t_i\sigma$.  Let $\mu$ be the most general unifier of $u$
    and $t_i$, then $\sigma=\theta\mu$, and we can simplify $\call{C}$
    by applying the first transformation rule \textsf{Unif},
    $\call{C}\Longrightarrow\call{C}'=\set{\call{C}_\alpha\mu,\call{C}_\beta\mu}$.
    We have $\sigma\models_{\mathcal{I}_\emptyset}\call{C}_\alpha$ and
    $\sigma\models_{\mathcal{I}_\emptyset}\call{C}_\beta$, then
    $\theta\models_{\mathcal{I}_\emptyset}\set{\call{C}_\alpha\mu,\call{C}_\beta\mu}.$
  \item If $t_i\sigma\notin (E_i\setminus\Variables)\sigma$ there
    exists a derivation starting from $(E_i\setminus\Variables)\sigma$
    of goal $t_i\sigma$, and then from $E_i\sigma$ of goal
    $t_i\sigma$.  By lemma \ref{lemma:00:saturation2}, there exists a
    derivation starting from $E_i\sigma$ of goal $t_i\sigma$ such that
    for all steps in the derivation such that $l\to r$ is the
    applied rule with the substitution $\sigma$, for all $s\in l$
    and $s\notin\Variables$, we have $s\sigma\subseteq E_i\sigma$.
    This implies that we can reduce $\call{C}$ to $\call{C}'$ by
    applying the \textsf{Apply} rule of transformation and
    $\theta\models_{\mathcal{I}_\emptyset}\call{C}'$.
  \end{itemize}
  We deduce that for all satisfiable constraint systems $\call{C}$ such that $\call{C}$ is not in solved
  form, $\call{C}$ can be reduced to another satisfiable constraint system $\call{C}'$ by applying the
  transformation rules. 
  When applying the transformation rules to a constraint system, we reduce its complexity (Lemmas
  \ref{lemma:00:terminaison1} and \ref{lemma:00:terminaison2}),  this implies that when we reduce $\call{C}$, we will  obtain at some step  a satisfiable constraint system which can not be reducible, this constraint system is in solved form. This concludes the proof.
  \end{proof}

\begin{lemma}{\label{Lemma:00:correction}}(Correctness.)
 Let $\call{C}=\set{(E_i \rhd{} t_i)_{i\in\set{1,\ldots,n}} }$ and 
$\call{C}'=\set{(E'_i \rhd{} t'_i)_{i\in\set{1,\ldots,n}}} $ such that $\call{C}'$
is obtained by applying the basic-narrowing on the terms of $\call{C}$.
For every substitution $\sigma'$ such that $\sigma'\models_{\call{I}_\emptyset}\call{C}'$, there  exists a substitution $\sigma$ such that $\sigma\models_{\I}\call{C}$.
\end{lemma}
\begin{proof}
We have $\call{C}=\set{(E_i \rhd{} t_i)_{i\in\set{1,\ldots,n}} }$,
$\call{C}\leadsto^*_{b.n}\call{C}'$ and
$\call{C}'=\set{(E'_i \rhd{} t'_i)_{i\in\set{1,\ldots,n}}}$.
Let $\theta$ be the composition of substitutions applied in the basic-narrowing derivation,
for all $i\in\set{1,\ldots,n}$ we have $\norm{E_i\theta}=E'_i$ and $\norm{t_i\theta}=t'_i$.
Let $\sigma'$ be a substitution such that $\sigma'\models_{\call{I}_\emptyset}\call{C}'$,  for all $i \in\set{1,\ldots,n}$ we have 
$t'_i\sigma'\in\rhclos{E'_i\sigma'}{\call{I}_\emptyset}$, this implies that 
for all $i \in\set{1,\ldots,n}$
$t'_i\sigma'\in\rhclos{E'_i\sigma'}{\II}$ (Corollary \ref{coro:cor2}),
and then, for all $i \in\set{1,\ldots,n}$
$t'_i\sigma'\in\rhclos{E'_i\sigma'}{\I}$
(Lemma \ref{lemma:00:saturation1}).
From the fact that $\norm{E_i\theta_n}=E'_i$ and $\norm{t_i\theta_n}=t'_i$ for all
$i\in\set{1,\ldots,n}$, we deduce that $t_i\theta\sigma'\in\rhclos{E_i\theta\sigma'}{\I}$
for all $i\in\set{1,\ldots,n}$ and 
this conclude the proof.
\end{proof}

\section{Conclusion}\label{sec:conclusion}

Besides the actual decidability result obtained in this paper, we
believe that the techniques developed to obtain this result, while
still at an early stage, are promising and of equal importance.
Several recent work~\cite{baudet-ccs2005,ANR-RTA07} have proposed
conditions on intruder systems ensuring the decidability of
reachability with respect to an active or passive intruder. In a
future work we plan to research whether the given conditions imply the
termination of the saturation procedure and the termination of the
symbolic resolution.

\bibliographystyle{plain}
\bibliography{constraint}

\end{document}